\documentclass{ws-procs9x6}

\begin{document}
\title{Measurement of the analyzing power of proton-carbon elastic 
scattering in the CNI region at RHIC}
\author{
\underline{O.~Jinnouchi$^{1}$}, 
I.G.~Alekseev$^2$, A.~Bravar$^3$, G.~Bunce$^{3,1}$, S.~Dhawan$^4$,  
H.~Huang$^3$, G.~Igo$^5$, V.P.~Kanavets$^2$, K.~Kurita$^{6,1}$, 
H.~Okada$^{7}$, N.~Saito$^{7,8,1}$, H.~Spinka$^{9}$, 
D.N.~Svirida$^2$, J.~Wood$^5$
}
%
\address{
  ${}^1$RIKEN BNL Research Center~ 
  ${}^2$ITEP~
  ${}^3$BNL~
  ${}^4$Yale University~
  ${}^5$UCLA~
  ${}^6$Rikkyo University~
  ${}^7$Kyoto University~
  ${}^8$RIKEN~
  ${}^9$ANL \\
}
\maketitle
\abstracts{
The single transverse spin asymmetry, ${\rm A_N}$, of the p-carbon 
elastic scattering process in the Coulomb Nuclear
Interference (CNI) region was 
measured using an ultra thin carbon target and polarized 
proton beam in the Relativistic Heavy Ion Collider (RHIC) at 
Brookhaven National Laboratory (BNL).
In 2004, data were collected to calibrate the p-carbon process 
at two RHIC energies (24~GeV, 100~GeV). 
${\rm A_N}$ was obtained as a function of momentum transfer $-t$. 
The results were fit with theoretical models 
which allow us to assess the contribution from a 
hadronic spin flip amplitude.
}
The elastic scattering of polarized protons off nuclei at 
RHIC energies $(24-250~{\rm GeV})$ 
provides information on a spin 
dependent hadronic spin-flip amplitude.
In the Coulomb-Nuclear 
Interference (CNI) region, i.e. 
$0.005 < -t < 0.05~{\rm (GeV/c)^2} $ where 
$t=(p_{out}-p_{in})^2 \approx -2M_CT_{kin}<0$, 
the single transverse spin asymmetry, $A_N$, of p-carbon elastic 
scattering is used to measure the beam 
polarization in RHIC. 
The data also provide physics information on the spin dependent 
hadronic contribution to the transverse asymmetry. 
For very small angle scattering, the elastic process dominates,
and experimentally the elastic events are cleanly identified 
by measuring the recoil carbons for polar angles near 
90${}^{\circ}$ in the laboratory frame.  
In the CNI region, the electromagnetic and hadronic helicity amplitudes 
are comparable in size.
A non-zero $A_N$ arises mainly from the interference between the coulomb 
spin-flip amplitude (which generates the anomalous 
magnetic moment of the proton) and the hadronic non spin-flip amplitude. 
This interference term, called 'pure CNI' is precisely 
determined from QED calculation. 
However, a contribution to $A_N$ can also come from 
the other interference term from the 
hadronic spin flip amplitude (coupling with 
the coulomb non spin-flip amplitude), 
which can be described
by Regge poles exchange phenomenology\cite{Larry}.
Since this hadronic term is in the nonperturbative QCD region, experimental 
data is indispensable.

Following the analogy to the helicity amplitude formalism of 
proton-proton elastic scattering, the $pC$ process can be described by two
amplitudes, spin non flip $F_{+0}(s,t)$, and spin flip $F_{-0}(s,t)$. 
The spin flip amplitude parameter $r_5^{pC}(t)$ is defined as,  
$r_5^{pC}(t) = {mF_{-0}^h}/(\sqrt{-t}~{\rm Im}F_{+0}^h)$,
where $m$ is the proton mass, and $F^h_{\pm0}$ is the hadronic
element of the helicity amplitudes.
This may be connected to a $t$ independent parameter  
$r_5$ for $pp$\cite{Kopeliovich}. 
The AGS experiment E950\cite{E950} has provided the only measurement of 
$A_N^{pC}(t)$, at $21.7~{\rm GeV/c}$. The $r_5$ result from E950 was
${\rm Re}~r_5 = 0.088\pm0.058~/~{\rm Im}~r_5 = -0.161\pm0.226$ 
with a strong anti correlation between real and imaginary parts. 

For the current experiment,
the polarized proton beam passes through an ultra-thin carbon 
ribbon target (3.5-$\mu g/{\rm cm^2}$ thick\cite{target}), 
and carbon recoils from CNI scattering are observed in six 
silicon strip detectors placed at 90${}^\circ$ to the beam direction, 
15~cm away from the target. 
Each detector has $10\times {\rm 24mm^2}$ active area 
divided into 12 strips, each directed parallel to the beam line. 
The six detectors are mounted inside the vacuum chamber with readout 
preamplifier boards directly attached to feed-through 
connectors on the detector ports. 
Data acquisition is based on waveform digitizer modules (WFD)\cite{WFD}. 
The system reads out the data without deadtime to accommodate very 
high event rates. The waveforms from the strips are digitized, and 
energy ($E$) and time of flight ($TOF$) w.r.t. the RHIC rf 
clock are determined 
by on-board FPGA. Typically $2\times10^7$ samples of 
carbon events are stored in memory on board and are readout 
after the measurement. 

Slow particles inside our kinematical acceptance follow the 
non-relativistic kinematics i.e., 
$TOF=\sqrt{\frac{M_CL^2}{2}}\frac{1}{\sqrt{E}}$,
which is slightly deformed by an energy loss correction
in the inactive silicon surface. The size of the deformation is used 
to estimate the thickness of the inactive layer\cite{run-03}.
It is estimated to be $57\pm12\mu {\rm g/cm^2}$. 
The invariant mass of the recoil particle is reconstructed with 
the time and energy information. A three standard deviation cut
around the carbon peak is applied for the carbon identification.
The cut clearly separates the carbon events ($11.17~{\rm GeV/c^2}$) 
from $\alpha$ background ($3.7 {\rm GeV/c^2}$). 

\begin{figure}[t]
\begin{center}
\includegraphics[scale=.26, angle=-90, keepaspectratio]{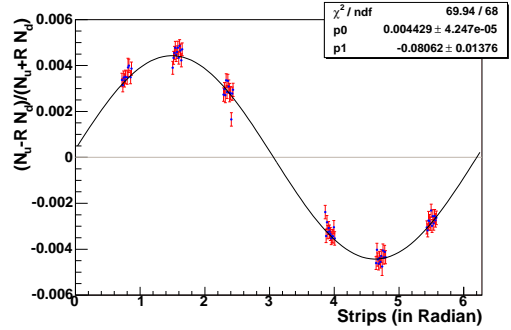}
\end{center}
\caption{Strip by strip asymmetry measurements. The raw up-down 
asymmetries are plotted as a function of strip location in radian.
The curve is the best fit with the sine function allowing phase shift, i.e.
$f(\phi) = P_0\sin(\phi+P_1)$}
\label{fig:stripbystrip}
\end{figure}

A raw asymmetry is calculated for carbon counts in left and 
right detectors using a geometric mean method, which takes advantage of the alternating spin patterns in RHIC\cite{SQROOT}.
Since each silicon strip can serve as an individual 
device measuring the asymmetry, the systematic uncertainty of the measurements
is estimated from the size of fluctuation among the strips.
Figure~\ref{fig:stripbystrip} shows the up-down asymmetry for the $i$-th
($i=1\cdots72$) strip ($\equiv(N^u_i-R\cdot 
N_i^d)/(N_i^d+R\cdot N_i^d)$ where $R$ is the luminosity ratio of up/down spin 
bunches). 
By allowing a phase shift 
to the $\sin\phi$ fit (two parameters: amplitude, phase), 
the $\chi^2/ndf$ is reduced to 70/68, whereas it is   
104/69 for one parameter fit. This $\chi^2/ndf$ value indicates 
there is a negligible systematic error in the measurement, and a
spin tilt ($4\sim5$ degrees) from vertical.
\begin{figure}[t]
\begin{center}
\includegraphics[scale=.30, angle=-90, keepaspectratio]{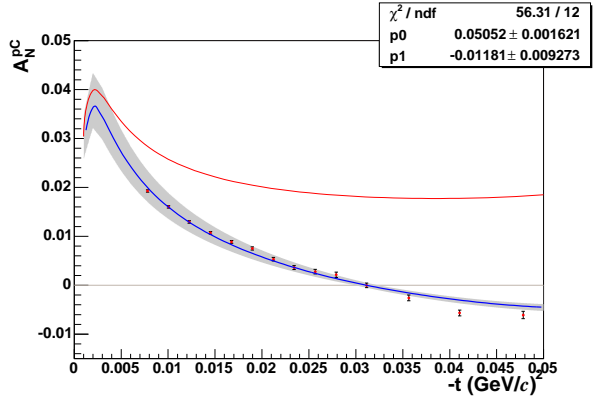}
\end{center}
\caption{Measured $A_N(t)$ at 100GeV. The calibration is carried out with
the polarization measured by the jet. The result is significantly different 
from the no hadronic spin flip calculation 
(the top curve). The shaded band represents 
the systematic uncertainty of the measurement}
\label{fig:A_N}
\end{figure}

$A_N^{pC}$ is obtained by dividing the raw asymmetry 
of $pC$ events by the beam polarization measured by
the polarized hydrogen gas jet target\cite{Jet},
Figure~\ref{fig:A_N} shows $A_N^{pC}(t)$ at 100GeV,
where $P_{beam} = 0.386\pm0.033$ \cite{Jet}.
The thin line through the data points is 
from a fit allowing hadronic spin flip\cite{Larry}.
The error bars on the data points are statistical only.
The size of systematic errors is shown with the shaded band. 
They are due to (i) the $-t$ ambiguity from the uncertainty
in the inactive surface layer of the silicon, (ii) the error 
of the beam polarization measured with the jet.  
The $r_5$ value from the best fit results is
${\rm Re}~r_5=0.051\pm0.002$ and ${\rm Im}~r_5=-0.012\pm0.009$.
The uncertainty is mainly due to the two systematic error sources
described above. The 1-$\sigma$ error contour has a very strong anti 
correlation between the real and imaginary part of $r_5$, from
$({\rm Re}~r_5, {\rm Im}~r_5) = $ (0.070, -0.16) to (0.035, 0.110).
For 24~GeV, even though the absolute scale of $A_N(t)$ is not available 
yet from the hydrogen jet measurement,
the shape is obtained and compared with 100~GeV for the range, 
$0.008<-t<0.028~{\rm (GeV/c)^2}$. The slope of $A_N^{pC}(t)$ 
at 100~GeV is significantly steeper than 24~GeV. 
%
\section*{Acknowledgments}
We are grateful to W.~Lozowski in Indiana University, Z.~Li and 
S.~Rescia in BNL for their hardware supports to this experiment.
The authors thank the jet target collaboration for providing the 
absolute beam polarization value. This work is supported by the U.S. 
Department of Energy and by RIKEN Laboratory, Japan.

\end{document}